\documentclass[twocolumn,showpacs,prb]{revtex4}

\usepackage{epsfig,wrapfig}
\usepackage{graphicx}
\usepackage{dcolumn}
\usepackage{bm}

\newcommand{\beq}{\begin{equation}}
\newcommand{\eeq}{\end{equation}}
\newcommand{\lsim}{\stackrel{\scriptstyle <}{\phantom{}_{\sim}}}

\newcommand{\sw}{\mbox{\scriptsize SW}}

\newcommand{\mn}{\mbox{\scriptsize min}}

\newcommand{\fl}{\mbox{\scriptsize FL}}

\begin{document}

\title{Properties of the Two-Dimensional Electron Gas Close to
the Fermi-Liquid Quantum Critical Point}
\author{
M.~Baldo,$^1$ V.~V.~Borisov,$^2$ J.~W.~Clark,$^3$
V.~A.~Khodel,$^{2}$ and M.~V.~Zverev$^2$  }
\vskip 0.8 cm
\affiliation {%
$^1$Istituto Nazionale di Fisica Nucleare, Sezione di Catania, \\
  I-95123, Catania, Italy \\
$^2$Russian Research Center Kurchatov Institute,
    Moscow, 123182 Russia \\
$^3$McDonnell Center
for the Space Science and Department of Physics, \\
Washington University, St.Louis, MO 63130, USA  } \vskip 0.3 cm

\vskip 0.5cm
\date{\today}

\begin{abstract}
The rearrangement of single-particle degrees of freedom of a dilute
two-dimensional electron gas in the vicinity of the quantum critical
point is examined within a microscopic approach.  It is shown
that just beyond the critical point, the Landau state undergoes
self-consistent rearrangement of the quasiparticle spectrum and
momentum distribution.  At very low temperatures, there emerges
a multi-connected quasiparticle momentum distribution.  With
increasing temperature, two crossovers occur: a fermion condensate
appears in the first and disappears in the second, giving way to
universal non-Fermi-liquid behavior.  Manifestations of these
crossovers in thermodynamic properties of the electron gas
are studied and characterized.  The four quasiparticle phases
predicted to exist in the vicinity of the critical point are
collected in a schematic phase diagram.
\end{abstract}


\maketitle

\section{ Introduction }

The two-dimensional (2D) electron gas is a widely used physical model
for the electron system of the inversion layer in high-quality
Si-MOS devices.  The density of that system can be varied over
several order of magnitude upon varying the external electric field
orthogonal to the layer.  The potential energy of the 2D
electron gas is proportional to the inverse of the average distance
between electrons, while its kinetic energy is proportional to the
inverse square of this distance.  Thus, with {\it decreasing} density,
the electron gas passes from a regime of weak correlations to one of
strong correlations.  Although experimental study of the 2D electron
gas began almost 40 years ago,\cite{Stiles-1968} it is only recently
that experiments \cite{Shashkin-PRB,Pudalov-PRL} have
reached a degree of accuracy such that
  the explosive growth of the effective mass $M^*$ of Landau
  quasiparticles is observed as the density approaches a critical
  value $\rho_{\infty}\simeq 8\times 10^{10}$\,cm$^{-2}$.
Evidence for
divergence of the quasiparticle effective mass is also found in 2D
liquid $^3$He (see Refs.~\onlinecite{Saunders-PRL,Godfrin-1,Godfrin-2})
and in some heavy-fermion materials.%
\cite{Coleman1,Stewart,Takahashi,Steglich,Custers,Gegenwart}
At zero temperature, the point $x=x_{\infty}$ of an external
parameter $x$ (density, pressure, magnetic field) at which the
effective mass is predicted or found to diverge is known as the
quantum critical point (QCP).\cite{Coleman1,Custers,Coleman-JPCM}
Properties of strongly correlated Fermi system in the vicinity
of the QCP exhibit non-Fermi liquid (NFL) behavior, i.e.\
they cannot be described within the Landau FL theory, as
ordinarily interpreted. It is often supposed
(see e.g.\ Refs.~\onlinecite{Hertz,Millis,Chubukov})
that the collective degrees of freedom are the {\it dramatis personae}
near the QCP; indeed, that they are responsible for the occurrence of
the QCP itself and for the properties of the system in its vicinity.
We follow a different strategy, exploring instead the possibility
that the single-particle (sp) degrees of freedom are the main characters
of the drama.  In this paper, we will show that the QCP is inherent in
Landau FL theory, and that a tool exists {\it within} this theory
for describing the system in the immediate vicinity of the QCP and
on {\it both} sides of it.

We focus on the 2D electron gas, for which a microscopic,
{\it ab initio} non-perturbative approach was developed in
  Refs.~\onlinecite{Shaginyan-SSC,physrep,bz}.
Applying this method, it was found
that at $T=0$ the effective mass $M^*$ of the 2D electron gas diverges
as the dimensionless parameter $r_s=\sqrt{2}Me^2/p_F$ (where $p_F$ is
the Fermi momentum) approaches a critical value $r_s^{\infty}\simeq 7$.
This result for the critical spacing parameter $r_s$ is
in reasonable agreement with the experimental value $\simeq 8-9$
cited in Refs.~\onlinecite{Shashkin-PRB,Okamoto-2004}.

Fig.~1 presents results \cite{bz} of evaluation of the single-particle
energy $\epsilon(p)$ (measured from the chemical potential $\mu$),
for the 2D electron gas at a set of $r_s$ values near
the critical point $r_s^{\infty}$.  As seen
in this figure, the spectrum acquires an inflection point at the critical
dilution, i.e., $\epsilon(p,r_s^{\infty})\propto (p-p_F)^3$, a situation first
analyzed in Ref.~\onlinecite{ckz}.  Beyond this point, meaning
$r_s>r_s^{\infty}$, the conventional Landau state, characterized at zero
temperature by a Fermi-step quasiparticle momentum distribution
$n_{\fl}(p)=\theta(p_F-p)$, becomes unstable against spontaneous creation of
quasiparticle-quasihole pairs. We call attention here to the fact that at $T=0$
a necessary condition for stability of the conventional Landau state is
positivity of the variation of the ground-state energy with respect to any
admissible variation of the quasiparticle momentum distribution from
$n_{\fl}(p)$.  More explicitly, it is necessary that \beq \delta E=\int
\epsilon(p,[n])\,\delta n(p) d\upsilon > 0 \label{necessary} \eeq holds for
$n(p) = n_{\fl}(p)$ and variations satisfying the particle-number constraint
\beq \int \delta n(p) \, d\upsilon = 0 \ , \label{dnorm} \eeq where
$d\upsilon=2d^2p/(2\pi)^2$. This stability condition (\ref{necessary}) is met
by $n_{\fl}$ provided the corresponding spectrum $\epsilon(p)$ has a single
zero at $p=p_F$. But as seen from Fig.~1,
  just beyond the critical point $r_s^{\infty}$ the spectrum takes on
    a cubic-like behavior with {\it three} zeros near the Fermi surface.
(It is worth noting that a similar shape for
the single-particle spectrum of 2D liquid $^3$He has been found
in the microscopic calculation of Ref.~\onlinecite{Krotscheck}
based on the method of correlated basis functions
\cite{Clark,Feenberg} (CBF).)  According to Eq.~(\ref{necessary}),
a spectrum $\epsilon(p)$ having this shape implies that one can
lower the energy of the system by introducing a quasihole at
a certain momentum $p<p_F$ for which $\epsilon(p)<0$, and a
quasiparticle at a certain momentum $p>p_F$ for which
$\epsilon(p)>0$.

In the general framework of Landau theory, the quasiparticle
spectrum is a functional
\beq
\epsilon(p,[n])={\delta E\over \delta n(p)}-\mu
\label{eps}
\eeq
of the quasiparticle momentum distribution $n(p)$; hence the
two quantities should be evaluated self-consistently.
The functional approach as formulated and applied in
Refs.~\onlinecite{Shaginyan-SSC,physrep,bz} cannot be used
in the 2D electron gas beyond the critical point $r_s^{\infty}$,
since this approach has so far only been developed for
states of Fermi systems describable in terms of standard
Landau quasiparticles having the Fermi-step momentum
distribution $n_{\fl}(p)$.  Moreover, extension of this
approach to finite temperatures remains incomplete.
\begin{figure}[t]
\includegraphics[width=0.7\linewidth,height=0.5\linewidth]{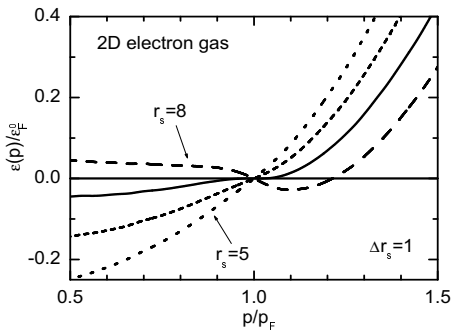}
\hskip 1.5 cm
\caption
{Single-particle spectrum $\epsilon(p)$ of the homogeneous
two-dimensional electron gas in units of $\varepsilon_F^0=p^2_F/2M$,
evaluated at $T=0$ for different values of $r_s$.
}
\label{fig:eg_sp}
\end{figure}

Nevertheless, an effective tool is available for self-consistent evaluation
of the single-particle spectrum $\epsilon(p)$ and momentum distribution
$n(p)$ beyond the point of instability of conventional Landau state,
and indeed within Landau theory itself. This tool is the basic
Landau formula
\beq
{\partial\epsilon(p)\over\partial {\bf p}} =
  {{\bf p}\over M} +
     \int\! f({\bf p},{\bf p'})\,
         {\partial n(p')\over\partial {\bf p'}}\, d\upsilon'
\label{landau}
\eeq
relating the group velocity of quasiparticles,
  $\partial\epsilon/\partial{\bf p}$,
to their momentum distribution
\beq
n(p)={1\over 1+e^{\epsilon(p)/T}}\, \quad (k_B=1) \,,
\label{fd}
\eeq
in terms of the Landau interaction function $f({\bf p},{\bf p'})$.
The quasiparticle interaction function is the second
variational derivative of the energy functional $E[n]$,
\beq
f({\bf p},{\bf p'},[n])={\delta^2 E\over \delta n({\bf p})\,
\delta n({\bf p'})} \ ,
\label{f_deriv}
\eeq
and hence itself depends on the quasiparticle momentum distribution
$n(p)$.  If non-zero temperatures are considered, the
$T$-dependence of this function should in principle also be taken
into account.  The theory presented here rests on two main
assumptions.  The first is that the difference between the interaction
$f({\bf p},{\bf p'},r_s,[n])$ corresponding to the {\it true}
quasiparticle distribution $n(p)$ at $r_s>r_s^{\infty}$, and
its counterpart $f({\bf p},{\bf p'},r_s,[n_{\fl}])$ evaluated with
Landau distribution $n_{\fl}(p)$, is small as long as the
relative size of the region of momentum space in which
$n_{\fl}(p)$ is altered, remains small.
Thus, we restrict our analysis to values of $r_s$ adjacent to
the QCP. The second assumption is that the $T$-dependent terms
in the quasiparticle interaction function produce contributions
to the single-particle spectrum of order of $T/\varepsilon_F^0$,
where $\varepsilon_F^0=p_F^2/2M$.  Thus, our analysis is
further restricted to the regime of very low temperatures
$T/\varepsilon_F^0\ll 1$.
These two assumptions allow us to use the quasiparticle interaction
function obtained from the functional approach in the
Landau relation (\ref{landau}), which becomes an equation for
self-consistent evaluation of the quasiparticle spectrum and
the quasiparticle momentum distribution.

In Sec.~II, we outline the approximation method employed for
evaluation of the quasiparticle interaction function of the
2D electron gas and the self-consistent scheme for subsequent
calculation of $\epsilon(p)$ and $n(p)$.  Working within this
framework, we then explore scenarios for rearrangement of
single-particle degrees of freedom of the system just beyond
its QCP (Sec.~III).  Section IV is devoted to calculation and
analysis of the corresponding thermodynamic properties of the
2D electron gas near the QCP.  In Section V, the quasiparticle
phase diagram of the system is constructed and discussed.
An Appendix presents further details on the calculation of
the quasiparticle interaction function.

\section{ Single-particle spectrum and quasiparticle interaction
function }

The well-known Feynman-Hellman formula, written as
\beq
E=\tau-{1\over 2} \int\limits_0^{e^2}\!\! de^2 \! \int\! dq
\int\limits_0^{\infty} {d\omega\over \pi}\,
[{\rm Im}\,\chi(q,\omega) +\pi\rho\delta(\omega)]
\label{ee}
\eeq
for the 2D electron gas, relates the energy $E$ of the system
to its response function $\chi(q,\omega)$.  This relation enables
one to evaluate both the single-particle spectrum at $T=0$,
by calculating the first derivative of Eq.~(\ref{ee}) as in
Ref.~\onlinecite{bz}, and the quasiparticle interaction
function, as described herein.  The latter  is given by
\beq
f({\bf p},{\bf p'})= -{1\over 2} \int\limits_0^{e^2}\!\! de^2 \!
\int {d^2q\over 2\pi\,q}
\int\limits_0^{\infty} {d\omega\over \pi}\,
{\rm Im} \left[{\delta^2\chi({\bf q},\omega)\over
\delta n({\bf p})\,\delta n({\bf p'})}\right]
\label{ff}
\eeq
in term of the second derivative of $\chi({\bf q},\omega)$
with respect to the occupation numbers.

If the Landau quasiparticle interaction function is available,
Eq.~(\ref{landau}) can be treated as an equation for $n(p)$.
Specifically, making use of Eq.~(\ref{fd}) the Landau relation
can be recast in the form
\beq
-{T\over n(p)[1-n(p)]}\,{\partial n(p)\over\partial {\bf p}} =
  {{\bf p}\over M} +
     \int\! f({\bf p},{\bf p'})\,
         {\partial n(p')\over\partial{\bf p'}}\, d\upsilon' \ ,
\label{equation}
\eeq
with
  $n(p)=-\int_p^{\infty}(d n(p')/d p')\, dp'$.

A microscopic prescription for evaluation of the quasiparticle interaction
function is given in the Appendix. Although full numerical realization of
this scheme presents a formidable challenge, calculation of harmonics of
the interaction function is tractable.  Based on analysis of different
contributions to $f({\bf p},{\bf p'},r_s,[n_{\fl}])$ and computation
of its zeroth and first harmonics at the Fermi surface as functions
of $r_s$, we have developed a method for approximate determination
of the quasiparticle interaction function.  To elucidate the essence
of this approximation, we note first that if the momenta ${\bf p}$
and ${\bf p'}$ lie exactly at the Fermi surface, the interaction function
depends only on the difference $q=|{\bf p}-{\bf p'}|$.  Analysis has
shown that the main contribution to this function has a strongly
pronounced peak at $q_0\simeq 1.97\,p_F$.  Since our considerations
are restricted to the immediate vicinity of the QCP, only momenta $p$
and $p'$ close to $p_F$ enter Eq.~(\ref{equation}).  Accordingly, we
approximate
  $F({\bf p},{\bf p'})=N_0 f({\bf p},{\bf p'})$, where $N_0=M/\pi$,
  in terms of the analytical form
\beq
  F(q)=-{\gamma(r_s)\over (q^2/q_0^2-1)^2+\beta^2} \,,
\label{approx}
\eeq
which by construction is peaked at $q_0$.
To choose the other parameters specifying
this function, we have made {\it ab initio} calculations of the
zeroth and first harmonics of the interaction function and their
derivatives with respect to $r_s$, and used the results, namely
$F_0(r_s^{\infty})=-1.3$, $F_1(r_s^{\infty})=1.0$, and
$(r_s^{\infty}/F_1(r_s^{\infty}))dF_1/dr_s=1.05$, as data
to be fit by adjusting $\gamma$ and $\beta$ in
Eq.~(\ref{approx}).  In this way these parameters were
determined to be
$\gamma(r_s)=0.08(1+1.05\, r_s/r_s^{\infty})$ and $\beta=0.135$.

With the interaction function dependent only on $|{\bf p}-{\bf p'}|$,
Eq.~(\ref{landau}) reduces to the relation
\beq
\epsilon(p) =
  \epsilon_0(p) +
     \int f(|{\bf p}-{\bf p'}|)\,
         n(p')\,d\upsilon' \ ,
\label{equa}
\eeq
in which $\epsilon_0(p)=p^2/2M-\mu$ is the bare sp spectrum.
Eq.~(\ref{equa}), together with Eq.~(\ref{fd}) and the
normalization condition
\beq
\int n(p)\,d\upsilon=\rho \,,
\label{norm}
\eeq
provide a set of equations for self-consistent evaluation of
the quasiparticle spectrum $\epsilon(p)$ and momentum
distribution $n(p)$. The chemical potential $\mu$, which enters
Eq.~(\ref{equa}) as a constant of integration, is also determined
self-consistently.

\section{Rearrangement of single-particle degrees of freedom}

We have solved Eq.~(\ref{equa}) for $r_s$ values situated in the vicinity
of $r_s^{\infty}$ and on both sides of the QCP.  To analyze the
nature of the solutions, we start with the results shown in Fig.~2
at $r_s<r_s^{\infty}$, i.e., for densities above that of the QCP.
As seen in the bottom panels of this figure, where the derivative
$d\epsilon/dp$ is plotted against momentum $p$, the effective mass
grows rapidly as $r_s$ approaches the QCP.  The ratio $M^*/M$
at low temperature is about 6 at $r_s=6.8$, reaches 15 at $r_s=6.9$,
and diverges at $r_s=7.0$.  We observe that near the QCP, the effective
mass $M^*(r_s)$ decreases with increasing temperature, and the
closer $r_s$ is to $r_s^{\infty}$,
the more pronounced is the decrease of $M^*(T)$.
The results shown in the bottom-right panel of Fig.~2 are
in accord with the behavior of the group velocity predicted
by the inflection-point model of the QCP proposed in
Ref.~\onlinecite{ckz}, in that $p_F/M^*(T)$ is found to
vary approximately as $T^{2/3}$.

\begin{figure}[t]
\includegraphics[width=0.9\linewidth,height=0.65\linewidth]{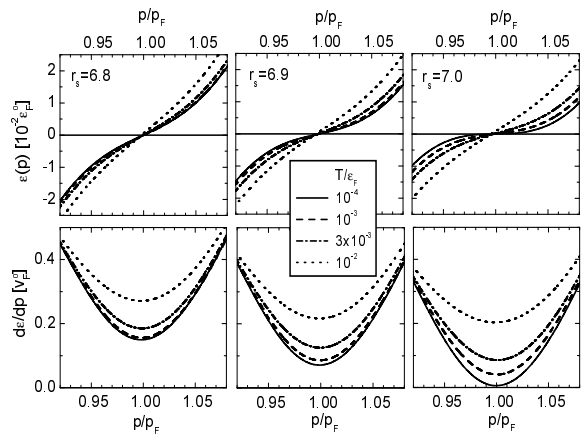}
\hskip 1.5 cm
\caption {Single-particle spectrum $\epsilon(p)$ in units of
$10^{-2}\,\varepsilon_F^0$ (top panels) and derivative
$d\epsilon(p)/dp$ in units of $v_F^0=p_F/M$ (bottom panels) for
the 2D electron gas at $r_s=6.8$ (left column), $r_s=6.9$ (middle
column), and $r_s=7.0$ (right column). The spectrum and its
derivative are shown as functions of $p/p_F$ at four (line-coded)
temperatures expressed in units of $\varepsilon_F^0$.}
\label{fig:eg_fl}
\end{figure}

Results from evaluation of the quasiparticle momentum distribution
$n(p)$ and spectrum $\epsilon(p)$ at $r_s>r_s^{\infty}$, i.e.,
for densities below that of the QCP, are displayed
in Figs.~3 and 4.  Inspection of Fig.~3 informs us that at low
temperatures in this region beyond the critical point, the quasiparticle
degrees of freedom are rearranged in a scenario involving
the formation of a multi-connected Fermi surface.  To the authors'
knowledge, Fr\"ohlich \cite{Frohlich} was the first to call
attention to the possibility of this type of Fermi surface
(see also Ref.~\onlinecite{Lifshitz}).
At $r_s=7.1$ and $r_s=7.2$, the rearranged quasiparticle momentum
distribution of the 2D electron gas at $T=0$,
for either spin projection,
is defined by a fully occupied circle ($n(p)=1$), surrounded
by an empty inner ring ($n(p)=0$), surrounded in turn by a fully
occupied outer ring, all centered on $p=0$.
At the small finite temperatures indicated in Fig.~3, increasing
$T$ leads to population of the inner ring ($n(p)>0$) at the expense of
depletion of the outer ring ($n(p)<1$).
The spectrum appears to have a cubic shape
\beq
\epsilon(p)\simeq D(r_s)\,{p_F(p-p_F)\over M} + \lambda
{(p-p_F)^3\over Mp_F} \,,
\label{cube}
\eeq
with three zeroes.  In this expression, the coefficient $D(r_s)$,
being negative beyond the QCP, vanishes at $r_s=r_s^{\infty}$,
while the coefficient $\lambda$ of the cubic term does not depend
on $r_s$.  As the system moves away from the QCP at $r_s > r_s^\infty$,
the relative size of the momentum region affected by the quasiparticle
rearrangement, measured by $\eta=(p_f{-}p_i)/p_F$, increases
as $\eta\propto (r_s/r_s^{\infty}{-}1)^{1/2}$.  With rising
temperature, the momentum distribution $n(p,T)$ and the ratio
$\epsilon(p,T)/T$ change dramatically.  At
$T/\varepsilon_F^0\sim 3\times 10^{-4}$ (for $r_s=7.1$) and
$T/\varepsilon_F^0\sim 10^{-3}$ (for $r_s=7.2$) the values
of $n(p)$ in inner and outer rings of the $T=0$ distribution
equalize, and the extrema in the spectrum $\epsilon(p)$
disappear.

\begin{figure}[t]
\includegraphics[width=0.9\linewidth,height=0.86\linewidth]{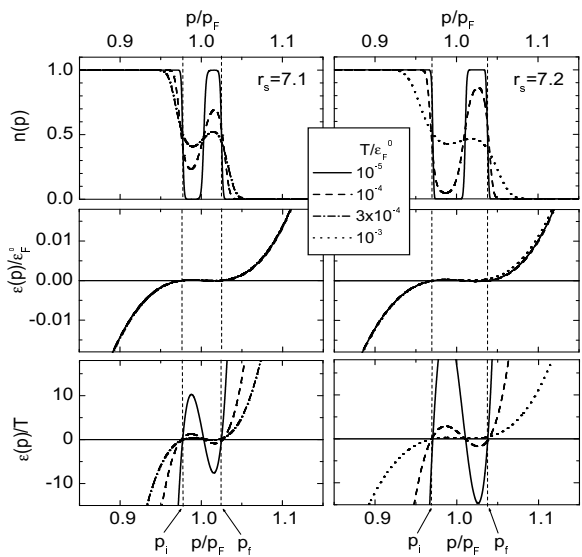}
\hskip 1.5 cm
\caption
{Occupation number $n(p)$ (top panels), single-particle spectrum
$\epsilon(p)$ in units of $\varepsilon_F^0$ (middle panels),
and ratio $\epsilon(p)/T$ (bottom panels) for the 2D electron gas
at $r_s=7.1$ (left column) and $r_s=7.2$ (right column). All three
quantities are shown as functions of $p/p_F$ at different temperatures
expressed in units of $\varepsilon_F^0$.}
\label{fig:eg_b}
\end{figure}

As the temperature continues to increase, the quasiparticle spectrum
and the momentum distribution each undergo a metamorphosis illustrated
in Fig.~4.  Inspection of the temperature dependence of these
quantities reveals a crossover from a state having a
multi-connected momentum distribution created by formation
of a ring-shaped ``bubble,'' to a state characterized by a region
of momentum space in which the ratio $\epsilon(p,T)/T$ and the
distribution $n(p)$ are independent of temperature.  The existence
of such a finite domain over which temperature independence
prevails is a signature of the phenomenon of fermion condensation,
i.e., the formation of a fermion condensate (FC) in the momentum
region involved.\cite{physrep,Nozieres}  Comparing Figs.~3 and 4,
it is seen that the size and location of the region of momentum
space occupied by the FC coincide respectively with
the size and location of the region in which
the ``bubble'' rearrangement  scenario -- depopulation of the
inner ring and population of the outer ring -- operates at lower
temperatures.

The emergence, at low but not the lowest temperatures, of a
sizeable domain in momentum space within which the occupation numbers
differ from 1 and 0, implies that there is a crossover from a
state with small entropy to another state with sufficiently large
entropy. To some extent the phenomenon that occurs is analogous
the Kosterliz-Thouless transition.\cite{Kosterliz-Thouless} Although
we are dealing with a crossover rather than a real phase transition,
this analogy enables us to estimate the temperature of the crossover.
To do so, we compare the free energies per particle $F=E-TS$ in the
bubble phase and the FC phase.  The entropy
associated with the bubble rearrangement is taken to be zero, while
that corresponding to the FC is estimated
   as
$\eta=(p_f{-}p_i)/p_F$.
  The loss in energy due to the transition from the bubble phase to
  the FC phase is estimated \cite{Baldo} as
$\Delta E\sim|D(r_s)|\,\eta^2\,\varepsilon_F^0,$ and the coefficient
of the linear term in Eq.~(\ref{cube}), as $|D(r_s)|\sim \eta^2$.
Thus, $\Delta E\sim \eta^4\varepsilon_F^0$. Since the difference
$\Delta S$ is of order $\eta$, the free energy of the FC state
becomes lower than that of the bubble state at
$T_0 \sim \Delta E/\Delta S\sim |D(r_s)|\,\eta\,\varepsilon_F^0\sim
|D(r_s)|^{3/2}\varepsilon_F^0$.

Proceeding to higher temperatures, the region belonging to the
FC gradually shrinks and finally disappears.  To analyze the type
of crossover taking place in this case, it is useful to study
thermodynamic characteristics of the 2D electron gas.

\begin{figure}[t]
\includegraphics[width=0.9\linewidth,height=0.86\linewidth]{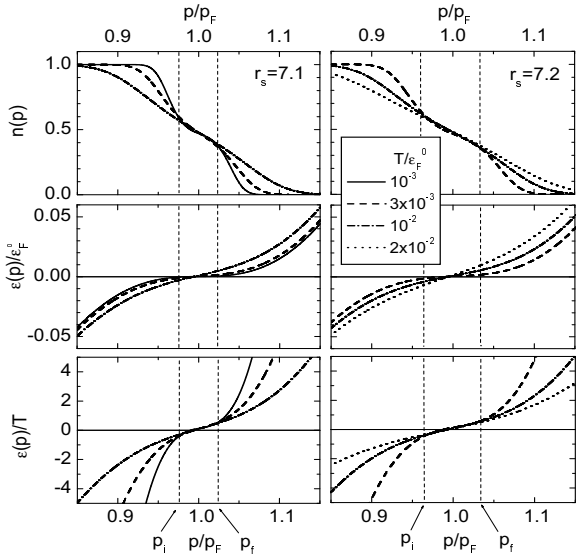}
\hskip 1.5 cm
\caption {Same as in Fig.~3 but at higher temperatures.}
\label{fig:eg_fc}
\end{figure}

\section{Low-temperature thermodynamic properties of 2D electron gas }

The density of states
\beq
N(T)=-\int {dn(\epsilon)\over d\epsilon} d\upsilon \label{dos}
\eeq
is a property well suited for analysis of the behavior of the 2D
electron gas in different temperature regimes.
The ratio of the inverse of this quantity to its inverse for the
corresponding 2D Fermi gas, $N_0^{-1}=\pi/M$, is plotted on a
log-log scale in Fig.~5. The horizontal line corresponds to FL behavior,
while the straight line with slope (or critical index) $\alpha$
corresponds to
the NFL regime with $N(T)\propto T^{-\alpha}$. As seen in the top
panel of Fig.~5, the FL behavior of the density of states at
$r_s<r_s^{\infty}$ persists only up to some characteristic temperature
$T^*$. In the region around $T^*$, a crossover occurs from FL
to NFL behavior, with a critical index $\alpha\simeq 0.6$.
This result is in essential agreement with the inflection-point model of
the QCP developed in Ref.~\onlinecite{ckz}, which yields $\alpha=2/3$.
The difference between 0.6 and 2/3 could be due to
  the term in the sp spectrum quadratic in $p-p_F$,
  which is absent in the inflection-point model.\cite{ckz}
According to the top panel of Fig.~5, $T^*\sim
10^{-3}\varepsilon_F^0$ at $r_s=6.8$. As $r_s$ approaches the
point $r_s^{\infty}=7.0$, the temperature $T^*$ decreases and
vanishes at the QCP.  This is apparent in the middle panel of Fig.~5:
at $r_s=7.0$ the inverse of the density of states, $N^{-1}(T)$,
  shows        universal NFL behavior with a critical index
$\alpha\simeq 0.6$ within the entire range of temperatures under
consideration.  As demonstrated in the bottom panel of this figure, at
$r_s>r_s^{\infty}$ the quantity $N^{-1}(T)$ undergoes two crossovers
separating three different temperature regimes. At low temperatures, it
exhibits FL behavior with a large effective mass.  In the vicinity
of $T_0$  (which is roughly $10^{-4}\varepsilon_F^0$ at $r_s=7.2$), a
crossover occurs from the FL regime to
  an intermediate NFL regime.
With further increase of temperature, another
crossover occurs in the vicinity of $T^*$ (which is roughly
$5 \times 10^{-3} \varepsilon_F^0$ at $r_s = 7.2$), leading to
the NFL regime with critical index $\alpha \simeq 0.6$.

\begin{figure}[t]
\includegraphics[width=0.54\linewidth,height=0.9\linewidth]{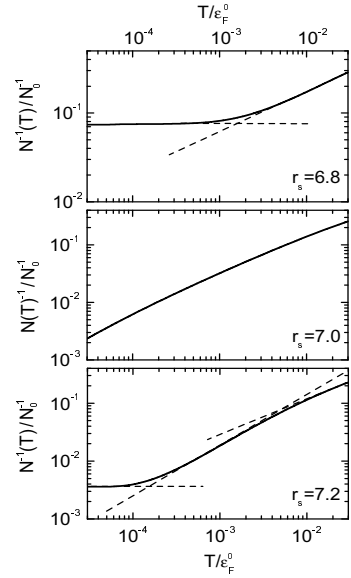}
\hskip 1.5 cm
\caption {Inverse of density of states $N^{-1}(T)$ in units of
$N_0^{-1}$ for the 2D electron gas at $r_s=6.8$ (top panel), $r_s=7.0$
(middle panel), and $r_s=7.2$ (bottom panel). Dashed lines show
different temperature regimes.}
\label{fig:eg_cross}
\end{figure}

Returning to Figs.~3 and 4, we may clarify the relationship of the three
temperature regimes exhibited by the density of states, to the
different regimes of behavior found in the quasiparticle momentum
distribution $n(p)$ and spectrum $\epsilon(p)$.  The bubble phase, with
its cubic-like spectrum (\ref{cube}), evidently corresponds to the FL
behavior of $N^{-1}(T)$. Smoothing of the bubble and flattening of
the quasiparticle spectrum $\epsilon(p)$ in the region $[p_i,p_f]$
corresponds to the crossover to the FC state. However, the
spectrum $\epsilon(p)$ is also flat in the regions adjacent to the
points $p_i$ and $p_f$, and hence these regions also play a
significant role in producing NFL behavior of thermodynamic
properties.  Indeed, the density of states,
\beq
N(T) \equiv {N_0\,[\nu_f(T)+\nu_n(T)]\over T} \ ,
\label{nu}
\eeq
decomposes into a sum of the FC term
\beq
\nu_f(T)= {1\over p_F} \int\limits_{p_i}^{p_f}
n_*(p)\,[1-n_*(p)]\,dp \sim \eta
\label{nu_f}
\eeq
and non-condensate part
\beq
\nu_n(T)\equiv \nu_n^>(T)+\nu_n^<(T)
\label{nu_n_}
\eeq
consisting of two terms.  Upon introducing
\beq
\nu_n(T;\epsilon_1,\epsilon_2) = {1\over
p_F}\int\limits_{\epsilon_1}^{\epsilon_2}
{n(\epsilon)\,[1-n(\epsilon)]\,d\epsilon \over \left(d\epsilon/
dp\right)} \ ,
\label{nu_nn}
\eeq
one can rewrite these two terms as $\nu_n^< =
\nu_n(T;-\mu,\epsilon_i)$ and
$\nu_n^>=\nu_n(T;\epsilon_f,\infty)$.  At $\eta\ll 1$
the group velocity near the boundary $p_f$ can be approximated
using
\beq
d\epsilon(p\to p_f)/dp=v_f(T)+v^{(2)}(p-p_f)^2 + \dots \ .
\label{groupn}
\eeq

\begin{figure}[t]
\includegraphics[width=0.85\linewidth,height=0.64\linewidth]{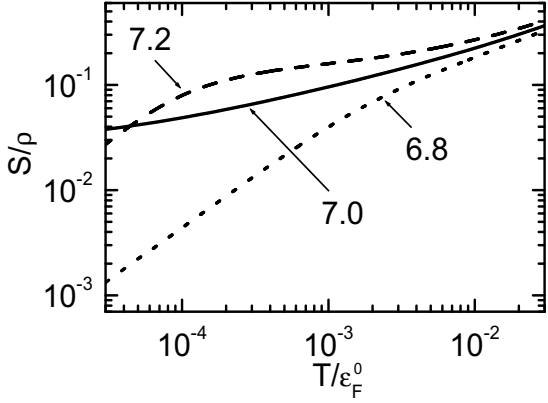}
\hskip 1.5 cm
\caption {Entropy per electron for the 2D electron gas as a function of
temperature in units of $\varepsilon_F^0$ at three values of
$r_s$.}
\label{fig:eg_entro}
\end{figure}

To estimate the value of $v_f(T)$, i.e.\ the slope of the FC
``plateau'' in the spectrum $\epsilon(p)$, we substitute the
momentum-independent result $n_*(p)$ for the momentum distribution
in the
region $[p_i,p_f]$ into the Fermi-Dirac formula (\ref{fd}) and obtain
\beq
\epsilon(p,T\geq T_0)=T\ln {1-n_*(p)\over n_*(p)} \ , \quad
p_i<p<p_f \ .
\label{spt}
\eeq
As seen in the bottom panels of Fig.~4, the width
$\epsilon(p_f)-\epsilon(p_i)\equiv \epsilon_f-\epsilon_i$ of the
FC zone at $T>T_0$ appears to be of order $T$ and almost
independent of $\eta$ for $ \eta_{\mn}\sim 10^{-2}$. Thus, at
$\eta>\eta_{\mn}$ the FC group velocity is estimated as
\beq \left({d\epsilon(p)\over dp}\right)_T
\sim {T \over \eta p_F} \ , \quad p_i<p<p_f \,,
\label{estfc}
\eeq
which in turn yields
$v_f(T)\sim T$.  The formula for the group
velocity outside the FC region but near the point $p_i$ is analogous
to (\ref{groupn}) with $v_i(T)\sim T$.  Using these results in
algebraic manipulations similar to those performed in
Ref.~\onlinecite{ckz}, we arrive at the relation
\beq
  N(T)=N_0\tau^{-1}[\nu_f    +\nu_n\tau^{1/3}]+{\rm const.} \,,
\label{sf}
\eeq
expressed in terms of the dimensionless temperature $\tau=T/\varepsilon^0_F$.

The FC is seen to play an insignificant role at small $\eta$,
say $\eta<\tau^{1/3}$, where the density of states behaves as
\beq
N(T)\propto T^{-2/3} \ .
\label{nfl1}
\eeq
In the opposite case, $\eta>\tau^{1/3}$, the FC contribution dominates in
Eq.~(\ref{sf}).  In our calculation, the intermediate situation
$\eta\sim\tau^{1/3}$ occurs, so
  the behavior of the density of states
in the region $T_0<T<T^*$ corresponds to a superposition of two terms
in Eq.~(\ref{sf}) of the same order.

The NFL difference
$\Delta\chi(T,\rho)=\chi(T,\rho)-\chi_{\fl}(\rho)$ between the magnetic
susceptibility and Pauli susceptibility is determined as
\beq
  \Delta\chi(T,\rho)\sim\mu_B^2N_0\tau^{-1}\,
  {\nu_f+\nu_n\tau^{1/3}\over
  1+G_0} \ ,
\label{dchi}
\eeq
where $G_0$, the zeroth harmonic of the spin-spin quasiparticle
interaction function, is independent of density in advance of
the QCP.\cite{Shashkin-PRB} The following behaviors of the magnetic
susceptibility derive from the above findings for the density of states.
At $\tau>\tau^*\propto\eta^3$, one has
\beq
\chi(T)\propto T^{-2/3} \,,
\label{chinfl1}
\eeq
whereas at $\tau<\tau^*$ the magnetic susceptibility of the
2D electron gas,
\beq
\chi(T)\propto T^{-1} \ ,
\label{curie}
\eeq
imitates that of a gas of localized spins.
At $\tau<\tau_0<\tau^*$ the Curie-like behavior
(\ref{curie}) is replaced by independence of $T$.

\begin{figure}[t]
\includegraphics[width=0.85\linewidth,height=0.64\linewidth]{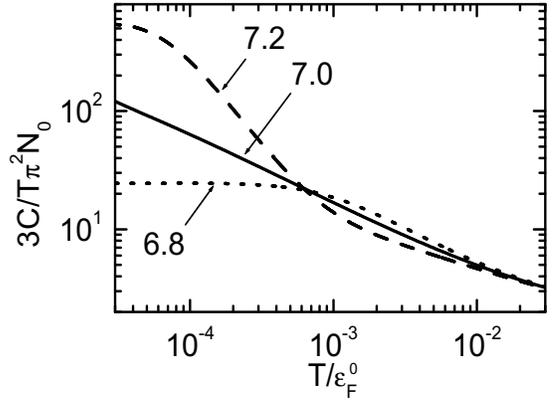}
\hskip 1.5 cm
\caption {Ratio $3C/T\pi^2N_0$ for 2D electron gas at three values
of the parameter $r_s$ is shown against temperature $T$ in units
of $\varepsilon_F^0$.}
\label{fig:eg_heatt}
\end{figure}

Fig.~6 shows the entropy per electron on a log-log scale at three values
of $r_s$.  Three different regimes may be discerned.  Two of them
are seen below the QCP, specifically at $r_s=6.8$, namely (i) the FL
domain with the usual linear temperature dependence and (ii) the NFL
regime where $S(T)\propto T^{1-\alpha}$ with $\alpha\simeq 0.6$.  At
the critical point, i.e., at $r_s=7.0$, the slope of the curve $S(T)$
is almost independent of temperature. Beyond the QCP, i.e.,
for $r_s > 7.0$, all three regimes are present, (i) the FL regime at
low $T$,
  (ii) the intermediate NFL regime corresponding the emergence of the FC,
and (iii) the NFL
regime with $\alpha\simeq 0.6$. Similar observations follow from
analysis of Fig.~7, where the ratio $C(T)/T$ of the specific heat
to temperature is depicted.  The sharp
drop of this ratio at $r_s=7.2$ corresponds to the crossover
to the FC state, where the temperature dependence of the entropy
  is weak.

\begin{figure}
\includegraphics[width=0.85\linewidth,height=0.64\linewidth]{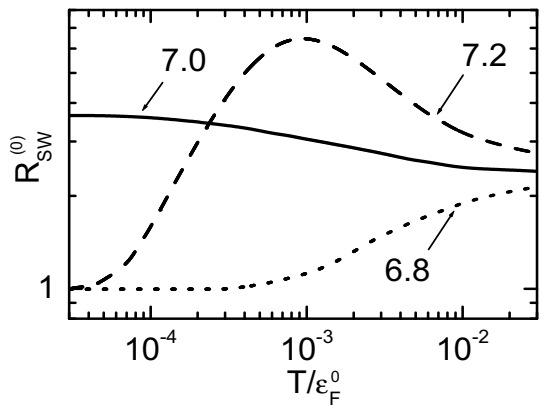}
\hskip 1.5 cm
\caption {Temperature dependence of the Sommerfield-Wilson ratio
$R^{(0)}_{\sw}=\pi^2T\chi_0(T)/3\mu_B^2C(T)$ for the 2D electron gas
in units of $\varepsilon_F^0$, plotted for three values of the
radius parameter $r_s$.}
\label{fig:eg_sw_ra}
\end{figure}

The evolution of the temperature regimes with passage through
the critical point are clearly revealed when the reduced Sommerfield-Wilson
ratio
\beq
  R^{(0)}_{\sw}(T)={\pi^2\chi_0(T)T \over 3\mu_B^2C(T)} \ ,
\label{sw_ratio}
\eeq
where $\chi_0(T)=(1+G_0)\chi(T)$,
is plotted against temperature for different $r_s$ values in
the critical region.  This quantity is unity for the FL, and is
usually strongly enhanced in case of NFL
behavior.\cite{Gegenwart-PRL-2005}  At $r_s=6.8$, the
crossover from FL to NFL behavior is reflected in an enhancement
of $R^{(0)}_{\sw}(T)$ for $T > 10^3 \varepsilon_F^0$. At $r_s=7.2$,
the ratio (\ref{sw_ratio}) goes to unity in the low-temperature limit,
confirming the FL character of the bubble phase, whereas its
enhancement to a value $\sim 7$ and subsequent decrease to $\sim
3$ with increasing temperature reflect the two crossovers
discussed above.

Up to this point, we have changed temperature at fixed values of $r_s$.
Let us now go across the critical value $r_s^{\infty}$ at fixed temperature.
The results for the density of states versus $r_s$ evaluated at different
temperatures are plotted in Fig.~9. Two different types of a behavior
of $N(r_s)$ can be distinguished.  Consider first the lowest of the
four curves, corresponding to $\tau=T/\varepsilon_F^0=10^{-3}$.  This
curve shows a monotonic increase of the density of states,
with a change of slope near the QCP, behavior which can be
understood as follows.  Below the QCP, i.e. at $r_s<r_s^{\infty}$,
the density of states $N(T,r_s)$ is proportional to the effective
mass $M^*(T,r_s)$ which, in the vicinity of the QCP,
behaves as \cite{ckz}
\beq
{M^*(r_s,T)\over M} \sim {1\over \kappa D(r_s)+ \nu\tau^{2/3}} \ .
\label{efm_qcp}
\eeq
Here, the positive constant $D(r_s)\propto 1-r_s/r_s^{\infty}$ is
the coefficient of the linear term (in $p-p_F$) of the
formula (\ref{cube}) for $\epsilon(p)$, which provides a good
approximation to the quasiparticle spectrum.
In Eq.~(\ref{efm_qcp}), $\nu=(9\lambda/4)^{1/3}$ is determined by
the $r_s$-independent parameter $\lambda$ controlling the cubic
term in Eq.~(\ref{cube}) (see Ref.~\onlinecite{ckz}), while
$\kappa$ is a constant of order one.
Beyond the QCP, a fermion condensate is present at
$T/\varepsilon_F^0=10^{-3}$.
The behavior of $N(r_s,T)$ in the state with a FC results
from the interplay of two contributions, as described by
Eq.~(\ref{sf}).
The first contribution, associated with the FC,
behaves like $\eta\tau^{-1}\propto |D(r_s)|^{1/2}\tau^{-1}$,
while the second, coming from regions adjacent to the FC
domain, is proportional to $\tau^{-2/3}$ and independent of $r_s$.

We turn now to the uppermost curve in Fig.~9, corresponding
to $\tau=3\times 10^{-5}$.  Below the QCP, the behavior
of this curve can also be understood in terms of Eq.~(\ref{efm_qcp}).
Beyond the QCP, the bubble rearrangement is favored by the low
temperature, and the resulting FL behavior of the density of states
can be again described by Eq.~(\ref{efm_qcp}), with all three
sheets of the Fermi surface taking part and making the
coefficient $\kappa$ considerably larger than unity.

An essential message of Fig.~9 is that at sufficiently low
temperature, the effective mass of the 2D electron gas extracted
from an analysis of its thermodynamic properties should
{\it decrease} beyond the QCP.

\begin{figure}[t]
\includegraphics[width=0.8\linewidth,height=0.75\linewidth]{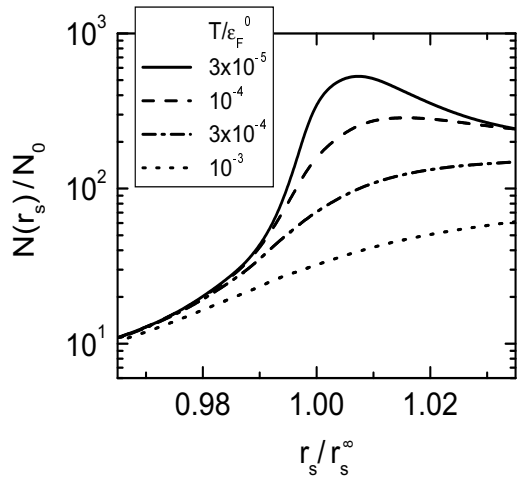}
\hskip 1.5 cm
\caption {Density of states $N(r_s)$ in units of $N_0$ for the 2D
electron gas at four values of temperature $T$ in units of
$\varepsilon_F^0$. }
\label{fig:eg_efm}
\end{figure}

\section{Quasiparticle phase diagram}

The information we have gained on the behavior of quasiparticle
degrees of freedom of the 2D electron gas near the quantum
critical point may be integrated into a quasiparticle phase diagram.
We do this with the caveat that our treatment has not taken into
account the feedback of the modification of single-particle
degrees of freedom on any pertinent collective modes.  Accordingly,
what we obtain is only a ``quasiparticle sketch'' of the real
phase diagram of the system.

Phase diagrams of other strongly correlated systems are customarily
drawn in a $T-x$ plane, where $x$ is some external parameters, notably
pressure, doping, or magnetic field.  Instead of these, we have
considered variations of the spacing parameter $r_s$ measuring the
extent to which the system is rarefied.  Thus, in Fig.~10 the phase
diagram of the 2D electron gas is drawn in $(T,x)$ variables,
with $x=1-r_s/r_s^{\infty}$.  On the top axis of the plot, we
show also the coefficient $D(x)\propto x$ of the linear term in
Eq.~(\ref{cube}) for the quasiparticle spectrum.  The
latter parameter is positive below the
QCP and negative above; its vanishing at the QCP reflects the
emergence of the horizontal inflection point in the quasiparticle
energy.  Beyond the QCP, the negative sign of $D(r_s)$ corresponds
to three roots of the equation
   $\epsilon(p)=0$,
and the presence of a bubble in the quasiparticle momentum
distribution.

\begin{figure}[t]
\includegraphics[width=0.88\linewidth,height=0.66\linewidth]{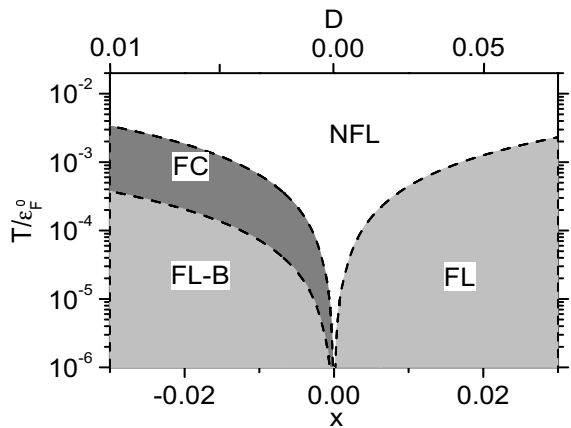}
\hskip 1.5 cm
\caption {Quasiparticle phase diagram of the 2D electron gas in $(T,x)$
variables with $x=1-r_s/r_s^{\infty}$. The dashed
curves show crossovers between the usual Fermi liquid (FL),
the Fermi liquid with bubble (FL-B), the fermion
condensate (FC) state, and the non-Fermi-liquid (NFL)
phase with critical index $\alpha\simeq 0.6$. The parameter $D(x)$
is indicated on the top axis.}
\label{fig:ph_di}
\end{figure}

Inspecting the phase diagram presented in Fig.~10, we start from
the side to the right of $x_\infty$, i.e., below the QCP,
where $D(r_s)>0$.  Here the low-temperature region of the diagram
is occupied by an ordinary Landau Fermi liquid.  The foregoing
analysis tells us that at a temperature around $T^*$, a
the crossover occurs from the FL regime to a universal NFL domain that
occupies the upper part of the diagram.  The dependence $T^*(D)$
can be estimated using Eq.~(\ref{efm_qcp}).  As a function of
$T$, the behavior of this expression changes at a temperature
for which $\nu\tau^{2/3}\sim D$.  Thus we estimate
that the temperature of the crossover to the NFL behavior in
the domain $x > 0$ goes like
\beq
T^*(D>0)\sim \nu^{-3/2}|D|^{3/2}\,\varepsilon_F^0 \ .
\label{tau_plus}
\eeq
This temperature is traced by the dashed line
in the right side of the phase diagram of Fig.~10, separating
the light gray FL region from the white NFL region.

Now consider the left side of the phase diagram lying
{\it beyond} the QCP, where $D(x)<0$ and two crossovers occur.
   Under increasing temperature there is first a crossover from
the FL state having a bubble (FL-B) to the FC state.  An estimate
of the temperature $T_0$ of this crossover was made in Sec.~III,
with the result $\tau_0\sim |D|\eta$. The dependence of $\eta$
on $D$ can be easily found by solving the equation $\epsilon(p)=0$,
which identifies the boundaries of the region in which the
Landau quasiparticle distribution is altered.  For the spectrum
as represented by Eq.~(\ref{cube}), this equation has the explicit
form
\beq
D\eta+\nu^3\eta^3/9=0 \ ,
\label{ddd}
\eeq
yielding
\beq
\eta=3 \nu^{-3/2}|D|^{1/2} \ .
\label{eta_d}
\eeq
Incorporating this dependence into $\tau_0 \sim |D| \eta$, we obtain
\beq
T_0(D<0)\sim 3\nu^{-3/2}|D|^{3/2}\,\varepsilon_F^0 \ .
\label{tau_0}
\eeq
The crossover in question is indicated in Fig.~10 by the curve in the
left part of the phase diagram separating the light-gray FL-B
regime from the gray region corresponding to the FC state.

The second crossover, from the FC regime to the universal NFL regime with
critical index  $0.6$,
occurs as the temperature increases up to the
value $T^*(D<0)$ at which the FC contribution to thermodynamic
characteristics of the system and that of regions adjacent to the
FC territory become comparable. For example, the FC contribution to the
density of states has been estimated in Sec.~IV as $\eta\tau^{-1}$, and
that from adjacent regions, as $\nu^{-1}\tau^{-2/3}$.  This implies that
$\tau^*(D<0)\sim \nu^3\eta^3$ and hence
\beq
T^*(D<0)\sim 27\,\nu^{-3/2}|D|^{3/2}\,\varepsilon_F^0 \ .
\label{tau_minus}
\eeq
The resulting dependence of $T^*(D<0)$ on $x$ is shown by the upper
curve in the left part of the phase diagram, which separates the FC
region from the NFL region.

As mentioned at the beginning of this section, the phase diagram
we have generated is a semi-finished product.  The next step
in building a complete phase diagram of the 2D electron gas
is to characterize the feedback from rearrangement of the
single-particle degrees of freedom on the collective modes.
It is expected that one of the Pomeranchuk stability conditions
is eventually violated, leading to collapse of the corresponding
collective mode.
Preliminary analysis
indicates that this violation may be due to unusual behavior of
the quasiparticle polarization operator beyond the QCP.
Future work will aim at determining which channel of
instability is the first to appear.

\section{Conclusions}

We have studied low-temperature properties of the 2D electron gas at
densities close to the point $\rho_{\infty}$ of divergence of
the quasiparticle effective mass.  A number of significant results
have been obtained by means of a microscopic approach based
on fundamental relations of the Landau Fermi-liquid theory and
an approximate evaluation of the quasiparticle interaction
function.

For $r_s<r_s^{\infty}$, i.e., for values of the dilution
parameter below the point of divergence of the effective mass,
we find that the temperature $T^*$ up to which the 2D electron
gas continues to exhibit FL thermodynamic properties,
decreases with increasing $r_s$ and vanishes as $r_s$ approaches
$r_s^{\infty}$.  At $T\sim T^*$, a crossover occurs to universal
NFL behavior of thermodynamic properties: the density
of states $N(T)$, the spin susceptibility $\chi(T)$, and the
ratio $C(T)/T$ of the specific heat to temperature scale universally
as $T^{-\alpha}$ with critical index $\alpha\simeq 0.6$.  This finding
is consistent with predictions of the inflection-point model
of the quantum critical point (QCP).\cite{ckz}

We have evaluated the spectrum $\epsilon(p)$ and the momentum
distribution $n(p)$ of quasiparticles beyond the QCP, i.e.\ at
$r_s>r_s^{\infty}$.
In this regime, the single-particle degrees of freedom are
altered, more or less profoundly, relative to the standard
FL picture.  At very low temperatures, the new quasiparticle
momentum distribution is multi-connected.  An empty ring
appears in the occupied Fermi circle, as the self-consistent
quasiparticle spectrum $\epsilon(p)$ acquires a cubic-like
shape with three zeroes at three boundary momenta.  This state,
however, possesses the usual Fermi liquid thermodynamic
properties, in that $N(T)$ is temperature independent,
as are $\chi(T)$ and the ratio $C(T)/T$.  Moreover, the
Sommerfield-Wilson ratio is unity, as it should be in a
FL state.

When the temperature rises to a certain value $T_0$, a
crossover to a state having a fermion condensate is found to
occur.
This more radical rearrangement of the single-particle degrees
of freedom is distinguished by {\it fractional} occupation of
momentum states, hence essential departure from the conventional
Landau Fermi-liquid picture, over a finite range of momenta
where the fermion condensate resides.  The momentum distribution
$n_*(p,T{>}T_0)$ in this domain, as well as the ratio
$\epsilon(p,T)/T$, proves to be independent of temperature;
the spectrum $\epsilon(p,T)$ itself, being proportional to $T$,
proves be almost flat.
The temperature dependence of the density of states $N(T)$,
and likewise for the spin susceptibility $\chi(T)$, is determined
by an interplay of a pure FC term proportional to $T^{-1}$ and a
NFL term proportional to $T^{-2/3}$.
The temperature dependence of the entropy is weak,
while the specific heat decreases sharply and the
Sommerfield-Wilson ratio jumps up to about 7.  With
further increase of temperature, a second crossover
is found, going from the FC state to the state with
universal NFL behavior.

Assembling results from the analysis of thermodynamic properties,
we have sketched a quasiparticle phase diagram of the dilute
2D electron gas in the immediate vicinity of the QCP.
There are four regions in this diagram: (i) a conventional
one corresponding the FL behavior of ordinary Landau
quasiparticles, (ii) the region associated with the multi-connected
quasiparticle distribution where FL temperature behavior
persists, (iii) the domain created by the FC rearrangement, and
(iv) the regime of universal NFL behavior.

Also of interest is our finding that at a fixed temperature
for which the multi-connected quasiparticle distribution is
favored, there is an enormous enhancement of the effective
mass $M^*$ as $r_s$ increases to the QCP, but no true divergence.
Moreover, beyond $r_s^{\infty}$, this enhancement is followed
by a decrease of $M^*$.  At temperatures for which the FC state
is favored, the rate of growth of the effective mass decreases
with increasing $r_s$
beyond the QCP.

New experiments using ultraclean Si-MOS samples could clarify the
nature of the QCP in 2D electronic systems inhabiting these structures.
\vskip 0.3 cm

The authors are grateful to V.~T.~Dolgo\-po\-lov, L.~A.~Maksimov,
Yu.~M.~Kagan, A.~A.~Shashkin, and G.~E.~Volovik for valuable
discussions. This research was supported by Grant
No.~NS-8756.2006.2 from the Russian Ministry of Education and
Science. MVZ thanks INFN (Sezione di Catania) for hospitality
during his stay in Catania.

\vskip 1 cm
\centerline{\bf APPENDIX}

\vskip 0.3cm

In this appendix we describe the scheme adopted for
evaluation of the quasiparticle interaction function
as expressed in Eq.~(\ref{ff}).  A microscopic functional
approach to this problem begins with the connection
 \beq
\chi(q,\omega)=\frac{\chi_0(q,\omega)}{1-R(q,\omega)\chi_0(q,\omega)}
\label{eff}
\eeq
of the linear response function $\chi(q,\omega)$ entering the
integrand in the r.h.s.\ of Eq.~(\ref{ff}) to the response function
$\chi_0(q,\omega)$ of the noninteracting gas, through the effective
interaction $R(q,\omega)$.  Accordingly, variation of the response
function $\chi(q,\omega)$ affects both the function
$\chi_0(q,\omega)$ and the effective interaction
$R(q,\omega)$:
$$
{\delta^2\chi(q,\omega)\over \delta n({\bf p})\,\delta n({\bf
p'})} = {\delta^2\chi_0(q,\omega)\over \delta n({\bf p})\,\delta
n({\bf p'})} \,\varphi^2(q,\omega)
$$
$$
+ 2\,\varphi^3(q,\omega)\,R(q,\omega)\, {\delta\chi_0(q,\omega)\over
\delta n({\bf p})}\, {\delta\chi_0(q,\omega)\over \delta n({\bf
p'})}
$$
\beq
+ \mbox{\rm terms containing}~{\delta R(q,\omega)\over
\delta n({\bf p})} \,,
\label{var2}
\eeq
where $\varphi(q,\omega)=[1-R(q,\omega)\chi_0(q,\omega)]^{-1}$.

The analysis performed in Ref.~\onlinecite{bz} indicates that the
shape of the effective interaction has a rather smooth dependence
on $r_s$.  For this reason, the terms containing the variational
derivative $\delta R(q,\omega)/\delta n({\bf p})$ serve only to
produce a slight renormalization of the chemical potential,
without contributing significantly to the shape of the spectrum.
By the same token, these terms are also neglected in the present
treatment.

We consider first the contribution $f^{(1)}({\bf p},{\bf p'})$
to the interaction function resulting from the product of
first variational derivatives of the function $\chi_0(q,\omega)$,
namely
\begin{widetext}
$$
f^{(1)}({\bf p},{\bf p'})=-2 \int\limits_0^{e^2}\!\! de^2 \!
\int {d^2q\over 2\pi\,q} \,R(q,\omega)
\biggl[ (1-n_{\fl}({\bf p}{-}{\bf q}))\,(1-n_{\fl}({\bf p'}{-}{\bf q}))\,
\delta(\epsilon_{\bf p}^0-\epsilon_{\bf p'}^0-
\epsilon_{{\bf p}{-}{\bf q}}^0+\epsilon_{{\bf p'}{-}{\bf q}}^0)
\,\varphi^3(q,\epsilon_{\bf p}^0{-}\epsilon_{{\bf p}{-}{\bf q}}^0)-
$$
$$
-\frac{(1-n_{\fl}({\bf p}{-}{\bf q}))\,
(\epsilon_{\bf p}^0{-}\epsilon_{{\bf p}{-}{\bf q}}^0)}
{(\epsilon_{\bf p}^0{-}\epsilon_{{\bf p}{-}{\bf q}}^0)^2
-(\epsilon_{\bf p'}^0{-}\epsilon_{{\bf p'}{-}{\bf q}}^0)^2}
\,\varphi^3(q,\epsilon_{\bf p'}^0{-}\epsilon_{{\bf p'}{-}{\bf q}}^0)
-\frac{(1-n_{\fl}({\bf p'}{-}{\bf q}))\,
(\epsilon_{\bf p'}^0{-}\epsilon_{{\bf p'}{-}{\bf q}}^0)}
{(\epsilon_{\bf p'}^0{-}\epsilon_{{\bf p'}{-}{\bf q}}^0)^2
-(\epsilon_{\bf p}^0{-}\epsilon_{{\bf p}{-}{\bf q}}^0)^2}
\,\varphi^3(q,\epsilon_{\bf p}^0{-}\epsilon_{{\bf p}{-}{\bf q}}^0)
+
$$
\beq
+\int\limits_{-\infty}^{\infty} {dw\over2\pi}
\frac{(\epsilon_{\bf p}^0{-}\epsilon_{{\bf p}{-}{\bf q}}^0)\,
(\epsilon_{\bf p'}^0{-}\epsilon_{{\bf p'}{-}{\bf q}}^0)}
{[(\epsilon_{\bf p}^0{-}\epsilon_{{\bf p}{-}{\bf q}}^0)^2{+}w^2]\,
[(\epsilon_{\bf p'}^0{-}\epsilon_{{\bf p'}{-}{\bf q}}^0)^2{+}w^2]}
\,\varphi^3(q,iw)
\biggr] \ .
\label{ff2}
\eeq
\end{widetext}
Following the strategy discussed in Section II, we assume that
both of the momenta ${\bf p}$ and ${\bf p'}$ are located near the
Fermi surface. We also assume both of them to be larger than $p_F$,
as is conventionally done in other problems of this kind.
Thus, we deal with the function $f^{(1)}(x)=f^{(1)}(p_F,p_F,x)$,
where $x = \cos({\bf p},{\bf p}')$,
and we are interested in the region $x\sim -1$ that corresponds
to backward scattering and provides the main contribution to the
shape of the spectrum near the Fermi surface.  It should be kept
in mind that within the diluteness regime under consideration, i.e.,
$r_s\sim r_s^{\infty}$, the function $\varphi(q,0)$ possesses
a sharp maximum at $q\simeq 2\,p_F$, such that the
the dominant contribution to the integral in Eq.~(\ref{ff2})
comes from the region where $q\sim 2\,p_F$.  Assuming
$\epsilon_{\bf p}^0=\varepsilon_{\bf p'}^0=0$, we note that
the first term of the integrand of Eq.~(\ref{ff2}) contains the
delta-function $\delta(\epsilon_{{\bf p}-{\bf q}}^0-
\epsilon_{{\bf p'}-{\bf q}}^0)$.  For $x\sim -1$, the condition
of equality of the energies $\epsilon_{{\bf p}-{\bf q}}^0$ and
$\epsilon_{{\bf p'}-{\bf q}}^0$ suppresses the integral over
angles of the vector ${\bf q}$, since it receives contributions
only from directions of ${\bf q}$ tangent to the Fermi surface
at the points ${\bf p}$ and ${\bf p'}$.  The first term of the
integrand in Eq.~(\ref{ff2}) yields appreciable
contributions to $f^{(1)}(x)$ only at $x\sim 1$, noting that
in this case all directions of ${\bf q}$ carry similar weight.
The second and third terms of the integrand contain expressions
$$
{\epsilon_{{\bf p}{-}{\bf q}}^0 \over
(\epsilon_{{\bf p}{-}{\bf q}}^0)^2%
{-}(\epsilon_{{\bf p'}-{\bf q}}^0)^2}
\qquad
\mbox{and}
\qquad
{\epsilon_{{\bf p'}{-}{\bf q}}^0 \over
(\epsilon_{{\bf p}{-}{\bf q}}^0)^2%
{-}(\epsilon_{{\bf p'}{-}{\bf q}}^0)^2}
\ ,
$$
respectively, thus implying that their contributions peak at
$\epsilon_{{\bf p}-{\bf q}}^0 \simeq \epsilon_{{\bf p'}-{\bf
q}}^0$. The same considerations as for the first term
show that neither of these terms contributes noticeably to
$f^{(1)}(x)$ at $x\sim -1$. In the fourth term, the region
$w\lsim 1$ dominates the integration over the variable $w$.
When evaluated within the approximation $R(q,w{\lsim}1)\simeq R(q,0)$
(which overestimates the integral), the fourth term produces a
contribution
\beq
\tilde f^{(1)}(x)\sim  -2 \int\limits_0^{e^2}\!\! de^2 \! \int
{d^2q\over 2\pi\,q} \, \, { R(q,0)\,\varphi^3(q,0) \over
\epsilon_{{\bf p}^{}_F{-}{\bf q}}^0 {+}\epsilon_{{\bf
p}^{\prime}_F{-}{\bf q}}^0}
\label{ff2t}
\eeq
to the interaction function.  An estimate of the corresponding
contribution to $f(x{\sim}-1)$, made by replacement of the
denominator in Eq.~(\ref{ff2t}) by its approximate value
$q^2/M$, indicates that the effect is rather small, namely
about $10 - 15\%$ at $r_s\sim 7$.  This contribution has not
been neglected, but instead is evaluated approximately.

We turn now to evaluation of the main term in the interaction
function, denoted $f^{(2)}({\bf p},{\bf p'})$, containing the
second variational derivative of the response $\chi_0(q,\omega)$.
This derivative is easily evaluated as
\beq
{1\over 2}\, {\delta^2\chi(q,\omega)\over 2\delta n({\bf p})\,
\delta n({\bf p'})}= (2\pi)^2\,
\delta(\epsilon_{\bf p}^0{-}\epsilon_{\bf p'}^0{-}\omega)\,
\delta({\bf q}{-}{\bf p}{+}{\bf p'}) \ ,
\label{varchi0}
\eeq
and we then obtain
\beq
f^{(2)}({\bf p},{\bf
p'})=-\int\limits_0^{e^2} de^2 {2\pi \over |{\bf p}{-}{\bf
p'}|}\, \varphi^2({\bf p}{-}{\bf
p'},\epsilon_{\bf p}^0{-}\epsilon_{\bf p'}^0) \ .
\label{f1}
\eeq
For momenta ${\bf p}$ and ${\bf p}'$ lying at the Fermi surface,
Eq.~(\ref{f1}) reduces to the expression
\beq
f^{(2)}(q)=-  {2\pi \over q} \int\limits_0^{e^2} de^2\,
\varphi^2(q,0) \ ,
\label{ffq}
\eeq
the integral over $e^2$ being evaluated numerically.

The interaction function $f^{(2)}(q)$ calculated for the 2D
electron gas at $r_s=6$, 7, and 8 is displayed in Fig.~11.
   The absolute value of this function
is seen to possesses a strongly pronounced maximum at
$q=q_c\simeq 2p_F$, which increases with increasing $r_s$.  The
peak and its enhancement with increasing $r_s$ are related to
the proximity of the system to an instability against spontaneous
appearance of charge-density waves (CDW) with wave vector
$q_c$ (see Refs.~\onlinecite{Pines-Nozieres,Swierkowski,Gold,bz}).
This situation is known to be responsible for divergence of the
effective mass.\cite{Shaginyan,Galitski}  In this regard, it is
worth noting that the {\it negative} sign obtained for the
quasiparticle interaction function $f(q{\sim}2p_F)$ provides
the correct positive sign of the first harmonic,
\beq
   F_1=N_0\int\limits_{-1}^1 f(x) { x\over \sqrt{1-x^2}} \,
       {dx\over 2\pi} \ ,
\eeq
  which equals 1 at the QCP.
At the same time, the
{\it negative} sign of the quasiparticle amplitude does not
support the idea that the quasiparticle weight $Z$ vanishes
near a CDW instability.  The latter eventuality is possible only
in case the quasiparticle amplitude, divergent at $q\sim 2p_F$,
has a {\it positive} sign, as erroneously assumed in
Ref.~\onlinecite{Chubukov} in considering enhancement of
the amplitude in the vicinity of such an instability.

\begin{figure}[t]
\includegraphics[width=0.72\linewidth,height=0.56\linewidth]{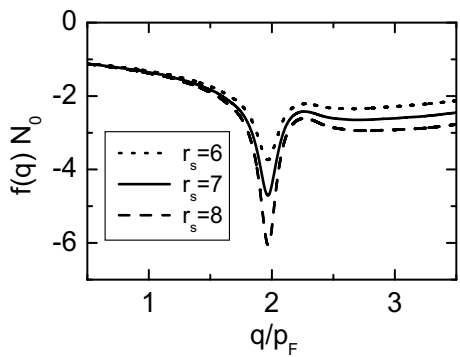}
\hskip 1.5 cm
\caption{Quasiparticle interaction function,
calculated within the microscopic functional approach for the 2D
electron gas and plotted as $ f(q)\,N_0$
at three values of the diluteness parameter $r_s$. }
\label{fig:eg_int}
\end{figure}

\end{document}